\documentclass[12pt]{article}
\usepackage{graphicx,amsmath}
\usepackage{units}

\newlength{\dinwidth}
\newlength{\dinmargin}
\renewcommand{\vec}[1]{\boldsymbol{#1}}

\def\lapproxeq{\lower .7ex\hbox{$\;\stackrel{\textstyle                                                    
<}{\sim}\;$}}                                                    
\def\gapproxeq{\lower .7ex\hbox{$\;\stackrel{\textstyle                                                    
>}{\sim}\;$}}                                                    
\def\be{\begin{equation}}                                                    
\def\ee{\end{equation}}                                                    
\def\bea{\begin{eqnarray}}                                                    
\def\eea{\end{eqnarray}}

\def\bb{\vec{b}'}

\def\bb{b\bar{b}}

\def\GeV{\rm GeV}

\def\sh{\hat s}
\def\sh2{{\hat s}^2}

\begin{document}

\begin{flushright}                                                    
IPPP/11/35  \\
DCPT/11/70 \\                                                    
\today \\                                                    
\end{flushright} 

\vspace*{0.5cm}

\begin{center}
{\Large \bf The small $x$ gluon and $b\bar{b}$ production at the LHC}

\vspace*{1cm}
                                                   
E.G. de Oliveira$^{a}$, A.D. Martin$^a$ and M.G. Ryskin$^{a,b}$  \\                                                    
                                                   
\vspace*{0.5cm}                                                    
$^a$ Institute for Particle Physics Phenomenology, University of Durham, Durham, DH1 3LE \\                                                   
$^b$ Petersburg Nuclear Physics Institute, Gatchina, St.~Petersburg, 188300, Russia \\          
                                                    
\vspace*{1cm}                                                    
                                                    
\begin{abstract}                                                    

We study open $\bb$ production at large rapidity at the LHC in an attempt to pin down the gluon distribution at very low $x$. For the LHC energy of 7 TeV, at next-to-leading order (NLO), there is a large factorization scale uncertainty. We show that the uncertainty can be greatly reduced if events are selected in which the transverse momenta of the two $B$-mesons balance each other to some accuracy, that is $|\vec p_{1T}+\vec p_{2T}|<k_0$. This 
will fix the scale $\mu_F\simeq k_0$, and will allow the LHCb experiment, in particular, to study the $x$-behaviour of gluon distribution down to $x\sim 10^{-5}$, at rather low scales, $\mu\sim 2$ GeV. We evaluate the expected cross sections using, for illustrative purposes, various recent sets of Parton Distribution Functions.

\end{abstract}                                                        
\vspace*{0.5cm}                                                    
                                                    
\end{center}

\section{Introduction}
The data from HERA and the Tevatron do not constrain the behaviour of the low $x$ gluon density, $g(x,Q^2)$. Indeed, if $Q^2 \sim 4~\GeV^2$, then already for $x\lapproxeq 10^{-3}$ there is a significant difference between the gluon distributions found in the different global PDF analyses. On the other hand, this is just the region sampled by the underlying events at the LHC.

It appears attractive to use the inclusive $b\bar b$ production at the LHC to study the behaviour of the gluon distribution in the very low-$x$  region. Indeed, due the rather large mass of the $b$ quark, the process may be described in the framework of perturbative QCD. The dominant contribution arises from the $gg\to b\bar b$ hard subprocess. Its cross section has the following structure
\be
\label{eq:sig}
d\sigma/d^3p~=~\int dx_1dx_2~g(x_1,\mu_F)~|{\cal M}(p;\mu_F,\mu_R)|^2~g(x_2,\mu_F)\ ,
\ee
where the gluon densities, $g(x_i,\mu_F)$, are taken at some factorization scale $\mu_F$, and the matrix element squared, $|{\cal M}|^2$, describes the cross section of the elementary $gg\to b\bar b$ subprocess. The process samples gluons which carry  momenta fraction $x_i$ of the initial protons, where
\be
\label{x12}
x_{1,2}=\frac{m_{\rm hard}}{\sqrt s}\exp(\pm y).
\ee
At the LHC energy of $\sqrt s=7$ TeV, and rather large rapidity\footnote{Rapidities in this range are optimal for the LHCb experiment \cite{LHCb}.}, $y\sim 5$, of the whole system produced in hard subprocess, one can probe the gluon densities with  $x\sim 10^{-5}$. For this estimate we have taken  the mass created in `hard
subprocess' $m_{\rm hard}=10$ GeV.
Recall that at present there are no data in this small $x$ domain and different global parton analysis predict quite different gluons, especially  close to the input scale for parton evolution, see, for example, \cite{MSTW08,CT10}. It therefore appears that the LHC, and the LHCb experiment in particular, offers a golden opportunity to make a precise determination of the gluon in this important low $x$ domain. However, first we must face the problem of the choice of factorization and renormalization scales.

A factorization scale $\mu_F$ is needed to separate the contributions hidden in the incoming Parton Distribution Functions (PDFs)  from those that included in the hard matrix element $|{\cal M}|^2$. Here, the gluon $g(x,\mu_F)$ is the PDF that we are concerned with. Contributions with 
low gluon virtuality $q^2<\mu_F^2$ are included in the PDF, while those with 
$q^2>\mu_F^2$ are assigned to the matrix element. 
The second scale, the renormalization scale $\mu_R$, in (\ref{eq:sig}) is necessary to fix the small value of QCD coupling, $\alpha_s(\mu_R)$, and to justify the perturbative QCD approach.
In principle, if all contributions (NLO, NNLO, etc.) are included, then calculated cross section would not depend on the values chosen for both of the scales $\mu_R$ and $\mu_F$.

\section{Problems associated with the choice of scales}
\begin{figure} [t]
\begin{center}
\includegraphics[height=8cm]{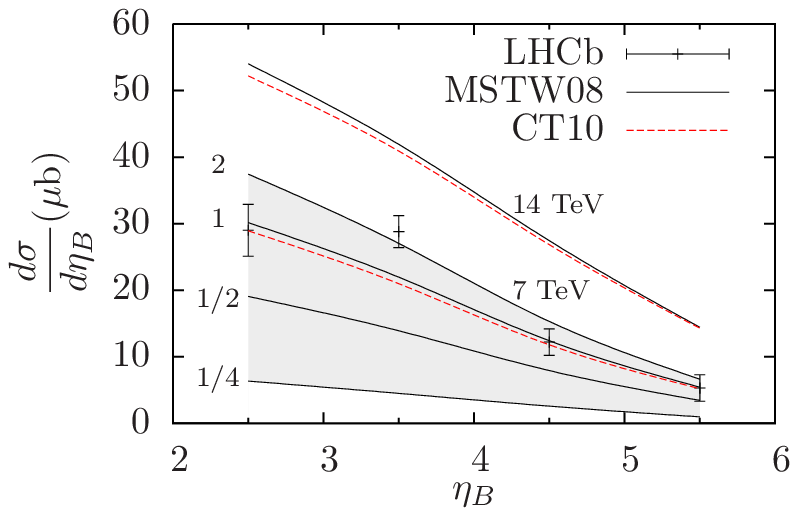}
\caption{\sf The NLO predictions for the cross section of $\bb$ production obtained using the FONLL program \cite{FONLL} from MSTW08 \cite{MSTW08} (continuous curves)  and CT10 \cite{CT10} (dashed curves) parton sets, at the LHC energies of 7 TeV and 14 TeV, as a function of pseudo-rapidity $\eta_B$ with scale $\mu_F=m_\perp$ and $m_b$=4.75 GeV; compared with LHCb data at 7 TeV \cite{LHCb}. The predictions using MSTW08 partons are also shown for four choices of factorization scale: $\mu_F=2m_\perp,~m_\perp, ~m_\perp/2, ~m_\perp/4$. The renormalization scale is set to $\mu_R=m_\perp.$}
\label{fig:A2}
\end{center}
\end{figure}

\begin{figure} [t]
\begin{center}
\includegraphics[height=8cm]{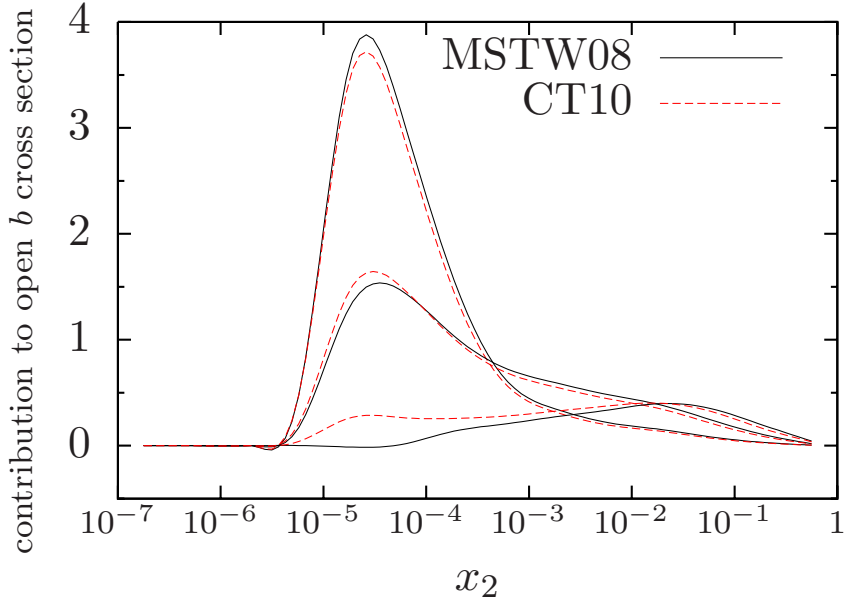}
\caption{\sf The distribution of the values of $x_2$ of the gluons sampled in NLO $\bb$ production with $\eta_b=5.5$ at the LHC energy of 7 TeV, after the cross section has been integrated over $x_1$ and $p_T$. The distributions are shown for the MSTW08 \cite{MSTW08} and CT10 \cite{CT10} parton sets for three values of the factorization scale, namely $\mu_F=m_\perp$ (upper pair of curves), $m_\perp/2$ and $m_\perp/4$ (lower pair of curves at small $x$). Note that there is a difference between $y$ and $\eta$, namely $\eta-y \simeq ~$ln$(m_\perp/p_T)$. This difference is accounted for in the FONLL calculation \cite{FONLL} used to produce this plot. The renormalisation scale is taken to be $\mu_R=m_\perp$.}
\label{fig:A1}
\end{center}
\end{figure}

However, one faces difficulties in the description of the new LHC data \cite{LHCb} for $\bb$ production.  We list these below. 
\begin{itemize}
\item The NLO  QCD prediction strongly depends on the choice of factorization scale, see Fig.~\ref{fig:A2}. For example, the result obtained with the choice $\mu_F=2m_\perp$ is more than twice larger than that for the case of $\mu_F=m_\perp/2$, where here
 $m_\perp\equiv\sqrt{p^2_T+m^2_b}$).
\item Moreover, at the NLO, we have a sizeable contribution from the $2\to 3$ ($gg\to b\bar b g$) subprocess, where one additional gluon is emitted in the hard collision.
This leads to a considerable smearing of the $x$ domain where we sample the incoming gluons. The smearing is especially strong if we adopt a low factorization scale, because then there is a large phase space allowed for gluon emission from the matrix element. Note that the probability of emission is enhanced by two large logarithms\footnote{The $\ln(1/x)$ enhancement is the main origin of the scale uncertainty observed in the collinear NLO approach at very small $x$. If we were to decrease the factorization scale $\mu_F$, then we have to move gluons with $p_{gT}\sim \mu_F$ from the PDF to the matrix element. The problem is that, at very low $x$, there may be several gluons emitted in the PDF, while only one gluon emission is allowed in the NLO matrix element. This spoils the compensation between the variations of $|{\cal M}|^2$ and the PDF, which should provide (and, indeed,  in the larger $x$ region, does provide) the stability of the results under scale variations.}: $\ln(m^2_\perp/\mu_F^2)$ and  $\ln(1/x)$. In particular, Fig.~\ref{fig:A1} shows that if we choose a
 scale $\mu_F=m_\perp/4$  then the major contribution comes from $x\sim 10^{-2}$, and not from
 $x\sim 10^{-5}$ as we had hoped. From this viewpoint it would be better to take a large $\mu_F$.
\item On the other hand, to differentiate between the low $x$ gluons it would be better 
to work with a relatively low $\mu_F$, where the difference between the different global PDF analyses is larger. At high scales $\mu_F$, a large fraction of low $x$ gluons comes from a region of much larger\footnote{Recall that a parton loses $x$ during DGLAP evolution.} $x$ in input
distribution, where 
the input distribution is already well constrained by  existing  data. Therefore, 
at  larger $\mu_F$, predictions for LHCb $b\bar b$ production, based on different PDF sets, become close to each other.
\item Moreover, recall that since for a large scale $\mu_F\sim 2m_\perp$ up to the half of the cross section originates from  rather heavy
virtual gluons\footnote{We have seen from Fig.~\ref{fig:A2} that $\sigma(\mu_F=2m_\perp)>2\sigma(\mu_F=m_\perp/2)$.}, the original perturbative
calculation, which assumes that the gluon virtuality $q^2$ is small in comparison with the quark mass (or $m_\perp$), becomes inconsistent.  
\item Finally, at NLO, we also have an unavoidable uncertainty in the prediction of $\bb$ production arising from the choice of the renormalization scale, $\mu_R$, 
which we will discuss in Section \ref{sec:muR}.
\end{itemize}
However, despite these difficulties, we show that it is possible to use the LHC $\bb$ data to make a measurement of the shape of the gluon PDF in the interval $10^{-5}\lapproxeq x \lapproxeq 10^{-2}$.

\section{Fixing $\mu_F$ by a cut on the vector sum of $p_T$'s}
 To overcome the problems associated with the choice of renormalization scale, we may ask for the measurement of the cross section of $b\bar b$ events in which the two quarks balance each other in the transverse momentum plane to some accuracy; that is ${\vec p}_{1T}\simeq - {\vec p}_{2T}$. In other words, to seek events which satisfy a cut on the vector sum of the transverse momenta of the heavy quarks, 
\be
\label{eq:cut-t}
|\vec p_{1T}+\vec p_{2T}|<k_0\ .
\ee
 Of course, in this way we will lose some of the cross section, but this should not be a problem with the available high LHC luminosity. Another point is that one cannot measure  the quark momentum directly. However the momentum of the $B$-meson can be measured\footnote{Even for the $B\to D\mu\nu$ decay, exploited in \cite{LHCb}, one can restore the full 4-momentum of the $B$-meson based on three constraints: we know the $B$-meson mass and its direction (two angles: $\phi$ and $\theta$), since the
position of the $B$-meson decay vertex is observed in the detector.}, and due to the strong leading effect\footnote{The $B$-meson carries more than 80\% \cite{petr,cn} of the original $b$ quark momentum. In this paper, we present results for $\bb$ production.} in $B$-meson production,  the event selection, proposed in (\ref{eq:cut-t}), can be performed with sufficient accuracy for those events with reasonably small transverse momenta of the $B$-mesons, say, $|\vec p_{TB}|<5$ GeV.
Moreover in order to better constrain the $x$ values of the gluons in our selected events, we may put an additional cut on the pseudo-rapidities of the $B$-mesons, say,
\be
\label{eq:cut-y}
|\eta_1-\eta_2|<1\ ~~~~ \mbox{or}\ ~~~~ \eta-0.5<\eta_1,\eta_2<\eta+0.5.
\ee 
Here, for illustration, we use the latter cut and present results as a function of $\eta$.

With the above kinematics it is natural to choose $\mu_F=k_0$. At first sight, in this way we appear to have excluded any gluon emission due to the NLO matrix element;
 a gluon with a transverse momentum, $p_{gT}$, less than $\mu_F$ 
 should be  included in the PDF, while one with $p_{gT}>\mu_F$ 
 spoils the cut (\ref{eq:cut-t}).
 However, this is not true at NLO. DGLAP evolution is written in 
 terms of parton virtualities $q^2=q^2_T/(1-z)$, where $z$ is the fraction 
 of parent parton momentum carried by the next (in this case, 
final) 
 parton. So, a relatively soft gluon
with $p_{gT}<\mu_F$ may correspond to $q^2>\mu^2_F$, and thus be assigned to the matrix element. However, this will happen mainly for  large $z$ close to 1, that is, in a situation where the emission of an additional
 (and now {\it soft}) gluon does not change the mass 
 $m_{\rm hard}$ 
 of the `hard block' too much; and thus  
does not smear out the low $x$ of the gluon sampled by the process.

\section{Procedure to calculate the effect of the proposed cuts}
To demonstrate how effective the proposed cuts will be in determining the gluon PDF at small $x$, we  
calculate the expected NLO cross section using two different recent sets of parton distributions, namely
MSTW08 \cite{MSTW08} and CT10 \cite{CT10}, for an LHC energy of $\sqrt{s}=7$ TeV.  To implement the cuts we adapt the subroutines for
the matrix elements squared of LO and NLO $\bb$ production that are given in the public MCFM program \cite{MCFM}. These subroutines use the expressions given in \cite{nason} for the NLO loop corrections to the $2\to 2$ subprocess, and the expressions given in \cite{ellis} for the
$2\to 3$ subprocess. Note that the infrared divergences arising from the emission of very soft gluons are regularized by the so-called `plus' prescription, which means that the singularities in the integrands, as the momentum fraction $z\to 0$, are tamed by
\be
\int_0^1dzf(z)\left[\frac 1z\right]_+=\int_0^1dz\frac{f(z)-f(0)}z .
\ee

The `plus' prescription is well justified at very low $z$, and corresponds to the Bloch-Nordsieck procedure for soft gluon radiation and for the inclusive cross section, where the $f(0)$ terms, added in the virtual loop and in the real NLO contributions, cancel each other. However, in the $2\to 3$ matrix element not only soft gluons are emitted. If $z$ is not small, the cancellation may be spoiled by our cuts. Therefore, to calculate the real $2\to 3$ contribution we
impose the restriction that the virtuality of any external (gluon or $b$-quark) line, after gluon emission, must be larger than the factorization scale $\mu_F^2$. All contributions with smaller virtualities are included either in the incoming parton distributions or in the quark fragmentation functions.  The remaining $2 \to 3$ contribution, with exactly the same kinematics, due to the additional $f(0)$ term, goes to the NLO loop correction to cancel  
the corresponding (unphysical, if $z$ is not too small) term in the loop correction to the $2\to 2$ subprocess\footnote{The contribution corresponding to $f(0)$ was calculated taking the matrix element for very soft gluon emission, keeping the momenta of incoming partons and the outgoing $b$-quark fixed. After this, the $1/z$ singular factor was replaced by the corresponding $1/z$ factor for the $2\to 3$ event which satisfies all of our proposed cuts.}
 coming from the `plus prescription'.

As default parameters we take the renormalization scale to be $\mu_R=m_b=4.75$ GeV,
 and the factorization scale to be
$\mu_F=2$ GeV. Also we take $k_0=2$ GeV in (\ref{eq:cut-t}) for the cut on the vector sum of the transverse momentum of the outgoing $b$ quarks.


\section{Factorization scale dependence \label{sec:muF} }

To illustrate the dependence of the predictions for $\bb$ production on the choice of the factorization scale, $\mu_F$, we evaluate the cross section for the production of $b$ and $\bar{b}$ quarks with both of their pseudo-rapidities in the interval $5<\eta_{1,2}<6$, first using  
$\mu_F=2$ GeV, and then for $\mu_F=4$ GeV. We repeat the exercise for the interval $2<\eta_{1,2}<3$. We use the 
CT10 NLO set of partons \cite{CT10}, which are available for 
$Q>Q_0=1.3$ GeV. For both choices of rapidity intervals, the cross section calculated with the higher scale, $\mu_F=4$ GeV, is 
about 3 - 4 times larger than that calculated with $\mu_F=2$ GeV.
Such a strong factorization scale dependence is due to the behaviour of the incoming parton densities. In the small $x$ domain, relevant for the LHC, the summation of the double logarithmic terms, 
\be
\Sigma_n c_n(\alpha_s\ln(1/x)\ln(\mu^2_F/Q^2_0))^n,
\ee
in the DGLAP evolution, leads to an 
\be
\exp\left(\sqrt{(4N_c\alpha_s/\pi)\ln(1/x)\ln(\mu^2_F/Q^2_0)}\right)
\ee
growth of the gluon density with increasing $\mu_F$. The exponential growth
comes from the sum over the possibilities of emitting different numbers of gluons.  The growth cannot be compensated by the `hard' matrix element, which at NLO level, allows for the emission of only one gluon.  This double-logarithmic effect is the main source of the strong factorization scale dependence of the predictions for the single $b$-quark inclusive cross section. 

On the other hand, if we choose a `large' value of the scale, $\mu_F>k_0$, then we invalidate our proposed 
`$p_T$' cut (\ref{eq:cut-t}). Recall that an {\it integrated}  parton density at a scale $\mu_F$
includes the effects of all partons with transverse momenta $k_t<\mu_F$; and the transverse momentum of an incoming parton with a `large' $k_t$ will spoil the 
$p_T$ balance in (\ref{eq:cut-t}). To control the transverse momenta of the incoming partons we may use {\it unintegrated} parton distributions\footnote{We also included the contributions from the incoming quarks,  which are, however, negligibly small in the low $x$ 
 region of interest.},
$f_g(x,k_t,\mu_F)$, and then integrate them over all $k_t<k_0$. These distributions can be obtained to NLO accuracy from the conventional integrated PDFs following the prescription of Ref.~\cite{MRW}. However, recall that integrated PDFs are not available at low scales, $k$, less than $Q_0$. Therefore we replace that part of the integral over the unintegrated PDFs with $k^2=k_t^2/(1-z)<Q_0^2$ by the `integrated' value $xg(x,Q_0)T(Q_0,\mu_F)$, where the Sudakov factor, $T$, accounts for the probability {\it not} to emit an extra parton, and thus not to enlarge $k_t$, during the evolution from $Q_0$ to $\mu_F$. The $T$ factor is given by
\be
T(Q_0,\mu_F)~=~{\rm exp}\left(-\int^{\mu_F}_{Q_0} \frac{d\kappa^2}{\kappa^2}\frac{\alpha_s(\kappa^2)}{2\pi}\int^1_0 dz~zP_{gg}^{\rm LO+NLO}(z)\right)
\ee
where the precise form of the splitting function $P_{gg}^{\rm LO+NLO}(z)$ is given in \cite{MRW}.

With such a procedure, for fixed $k_0$, the main double-logarithmic 
 effects are
 correctly included via the unintegrated PDF, 
  $f_g$, while the single-log dependence of  
$f_g$ on $\mu_F$ is mainly compensated by the NLO loop corrections in the `hard subprocess' cross section. 
The net effect of this procedure, based on `unintegrated PDFs', is a great reduction in the dependence on the choice of the factorization scale. For example, changing the scale $\mu_F$ from 2 to 4 GeV now leads to less than 20(35)\%
decrease in the prediction of the $\bb$ cross section in the intervals $2<\eta<3$ (and $5<\eta<6$), rather than the factor of 4 (3) increase, see Fig. \ref{fig:bb4} which we will introduce below.
Note that within a 30\% accuracy, the result obtained using the unintegrated PDFs coincides with that calculated in the conventional
NLO collinear framework with $\mu_F=k_0$.

\section{Renormalization scale dependence  \label{sec:muR}}

The dependence of the cross section on the choice of renormalization scale, $\mu_R$, arises from the running QCD coupling, $\alpha_s(\mu_R)$. In general, the variation of the QCD coupling should be compensated by the logarithmic terms, $\ln(\mu_R/\mu_F)$, in the NLO (and higher  order) virtual loop corrections \cite{smith}. Unfortunately, in the region of interest, say, $\mu_F=2$ GeV and $\mu_R\simeq (1 - 2)m_b$, the corresponding NLO contribution, that is the factor $\alpha_s^3\ln(\mu_R/\mu_F)$, is practically constant, and does not compensate the variation of the
LO term, $|{\cal M}^{\rm LO}|^2\propto \alpha_s^2(\mu_R)$. Therefore the result
calculated with $\mu_R=2m_b$ turns out to be about 30\% smaller than that with $\mu_R=m_b$.
Taken together with the uncertainty in the value of the mass, $m_b$, of the $b$-quark, 
this leads to an unavoidable  uncertainty ($\sim 50\%$), at NLO level, in the normalization of the $b\bar b$ cross section\footnote{For our computations, we choose the conventional value, $\mu_R=m_b=4.75$ GeV \cite{cn,jlz} which was used to describe $\bb$ production at the Tevatron.}.

Recall that at these scales the NLO corrections are rather large, larger than the size of the LO contribution. Thus it is not evident that the much more complicated NNLO calculation will improve
the accuracy significantly.

\section{Results for $\bb$ production after cuts}

Nevertheless, in spite of the uncertainties in the 
  normalization that we discussed above, the expected 
{\it ratio} of 
the cross sections measured in different rapidity intervals
is quite stable and is driven entirely by the $x$ behaviour of the gluon. We illustrate below how this enables LHC $\bb$ data to determine the {\it shape} of the gluon PDF in the interval  $10^{-5}\lapproxeq x \lapproxeq 10^{-3}$. 

Recall that after imposing the cuts of (\ref{eq:cut-t}) and (\ref{eq:cut-y}), the variation of the mass, $m_{\rm hard}$, created in the hard subprocess, is strongly limited. The contribution of the $2\to 3$ subprocess
 never exceeds 40\% of the whole cross 
section; typically it only amounts to about 1/3. Moreover, this $2\to 3$ contribution arises from relatively soft gluon emission, which does not change $m_{\rm hard}$ very much. An important consequence is that these $b\bar b$ events in different intervals of pseudo-rapidity, $\eta$, sample the gluon in rather narrow intervals of $x$, which allows a precise study of the shape of gluon $x$ distribution.

\begin{figure} [t]
\begin{center}
\includegraphics[height=6cm]{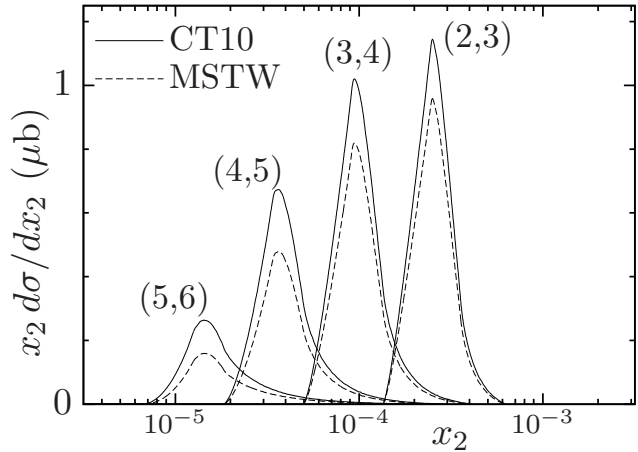}
\caption{\sf The $\bb$ cross section as a function of the momentum fraction of the proton carried by the slowest gluon, after cuts (\ref{eq:cut-t}) and (\ref{eq:cut-y}) have been imposed, for four different pseoudo-rapidity intervals taking $\eta=2.5,~3.5,~4.5,~5.5$ in (\ref{eq:cut-y}). In each case we show the predictions obtained using the NLO MSTW08 \cite{MSTW08} and CT10 \cite{CT10} parton sets. }
\label{fig:bb3}
\end{center}
\end{figure}
We illustrate this in Fig.~\ref{fig:bb3}.  It shows the distributions of the momentum fraction $x_2$ (where $x_2<x_1$) carried by the gluon in $\bb$ events if cuts (\ref{eq:cut-t}) and (\ref{eq:cut-y}) are imposed. The predictions are shown for four different intervals of the pseudo-rapidity, $\eta=2.5,~3.5,~4.5,~5.5$ in (\ref{eq:cut-y}), corresponding to the $b$ and $\bar{b}$ quarks both having rapidities in the intervals  (2,3), (3,4), (4,5) and (5,6) respectively. 
In each case we show the results obtained using MSTW08 and CT10 parton sets. 
We take $\mu_F=2$ GeV and $k_0=2$ GeV.  We see, for instance, that the $\bb$ events, selected by the rapidity cut (3,4), sample the gluon in quite a narrow range about $x=10^{-4}$.

\begin{figure} [t]
\begin{center}
\includegraphics[height=6cm]{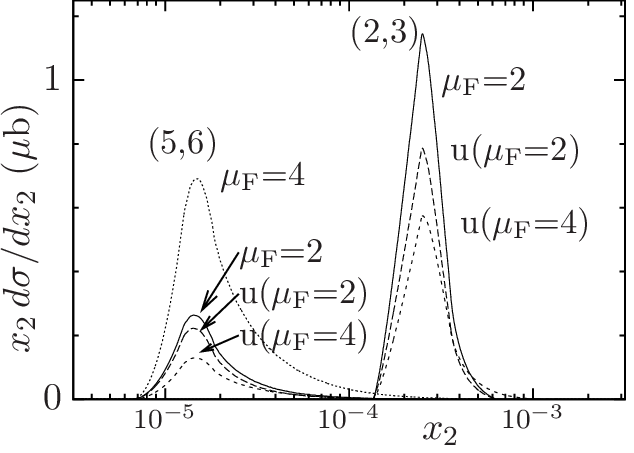}
\caption{\sf  The $\bb$ cross sections as a function of the momentum fraction carried by the slowest gluon, after cuts (\ref{eq:cut-t}) and (\ref{eq:cut-y}) have been imposed, predicted using integrated and unintegrated PDFs for two choices of factorization scale, $\mu_F=2$ and 4 GeV. The prefix ``u'' indicates unintegrated PDFs are used. The left and right plots correspond to the rapidity intervals (5,6) and (2,3) specified by taking $\eta=5.5$ and $\eta=2.5$ in (\ref{eq:cut-y}).  The $\mu_F=4$ GeV prediction obtained from integrated partons is not shown for the latter interval since it about 4 times higher than that for $\mu_F=2$ GeV. CT10 NLO partons \cite{CT10} are used.}
\label{fig:bb4}
\end{center}
\end{figure}

Fig.~\ref{fig:bb3} is obtained using integrated conventional PDFs. Fig.~\ref{fig:bb4} shows the effect of using the procedure based on unintegrated PDFs (as introduced in Section \ref{sec:muF}). The figure, obtained via CT10 partons, illustrates two effects. First, it shows the reduction in sensitivity to a change in factorization scale (from $\mu_F=2$ to 4 GeV) if unintegrated partons are used. Second, it shows that for our default choice, $\mu_F=2$ GeV, the predictions obtained using integrated and unintegrated PDFs are reasonably similar.

 We have emphasized that, despite 
  normalization uncertainties, the expected 
{\it ratio} of 
the cross sections measured in, say, the pseudo-rapidity intervals (2,3) and (5,6)
is quite stable and is driven entirely by the $x$ behaviour of the gluon. 
 If the gluon had a pure power behavior, $xg(x,\mu_F)\propto 
x^{-\lambda}$, then we would observe a {\it flat} $\eta$ dependence. This would follow since
\be
x_{1,2}\simeq \frac{m_{\rm hard}}{\sqrt s}\exp(\pm\eta) ~~~~~{\rm giving}~~~~~ x_1g(x_1)~x_2g(x_2)={\rm constant}.
\ee
 A 
non-trivial, that is non-flat, $\eta$ behaviour of the cross section will reflect the curvature (or a deviation from the power law) of the 
$x$ dependence of the gluon, and may be used to distinguish between the different sets of `global' PDFs.

 For example, in the case of CT10 integrated/conventional  partons the expected ratio for $\mu_F=2$ GeV is 2.86 ($\pm$ 15\% if we use the unintegrated PDF with $\mu_F=2$ or 4 GeV), while for the case of integrated MSTW08 partons the analogous ratio is 3.95.
 
\begin{figure} [t]
\begin{center}
\includegraphics[width=10cm]{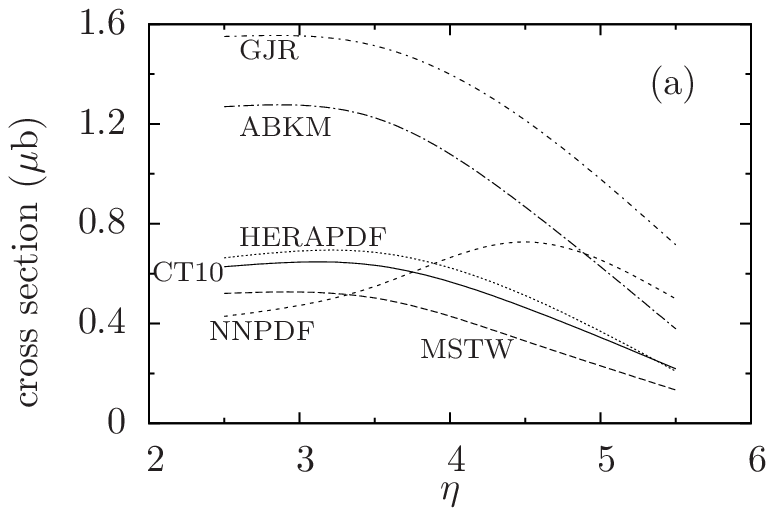}
\includegraphics[width=10cm]{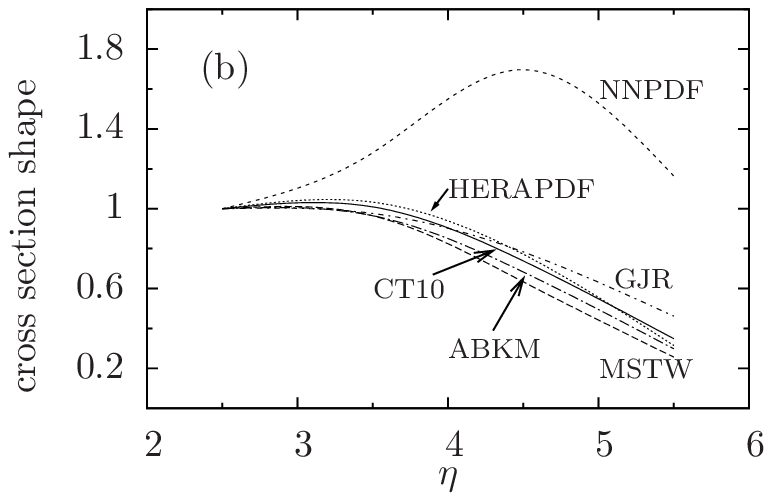}
\caption{\sf (a) The $\eta$ dependence of the $\bb$ cross section, after the cuts (\ref{eq:cut-t}) and (\ref{eq:cut-y}) have been imposed, obtained using six different NLO sets of partons MSTW08, CT10, NNPDF21, ABKM09-4, HERAPDF01 and GJR08VF \cite{MSTW08, CT10,PDF}; (b) the predictions normalized to one at $\eta=2.5$.}
\label{fig:bb5}
\end{center}
\end{figure}

Therefore a study of the $\eta$ dependence of the $\bb$ events, selected by the cuts, is a valuable way to study the small $x$ behaviour of the gluon.
In Fig.~\ref{fig:bb5}(a) we present the $\eta$ dependence of the cross section expected for six different NLO sets of PDFs, to demonstrate the sensitivity of such a method to the small $x$ behaviour of gluons. For this plot, $x_2$ varies from $x_2 \sim 3 \times 10^{-4}$ (corresponding to  $\eta = 2.5$) to $x_2 \sim 2 \times 10^{-5}$ (corresponding to  $\eta = 5.5$). This corresponds to the pseudo-rapidity range relevant to the LHCb experiment. In this small $x$ region the PDFs are unconstrained by existing data.  In particular, we note that the gluon distribution is well determined at $Q^2=5 ~\GeV^2$ to within 20$\%$ (typically less than 10$\%$) {\it only} if $x$ is in the range $10^{-3}\lapproxeq x\lapproxeq 0.3$, see Fig. 16 of \cite{MSTW08}. Since the gluons are unconstrained at very low $x$, we should anticipate the large spread of the predictions, shown in Fig.~\ref{fig:bb5}(a), for the $\eta$ dependence of the cross section after the cuts are imposed.  Indeed, if the errors on the PDFs were to be included, then the various predictions would overlap.

Recall that there is an overall normalization scale uncertainty of about 50$\%$ (see Section \ref{sec:muR}).
Even allowing for this, the observed $\bb$ cross section, $d\sigma/d\eta$, after cuts, will be provide valuable information about the gluon in this completely unexplored low $x$ domain. However, the normalization uncertainty does not affect the shape of the prediction of the cross section versus $\eta$, so the $\bb$ data are capable of yielding even more precise information.  Simply for illustration, we compare in Fig.~\ref{fig:bb5}(b) the shapes of the predictions of all the parton sets after they have been normalized to 1 at $\eta=2.5$.  Of course, the present huge PDF errors mean such a comparison cannot distiguish between them. Rather the $\bb$ data will determine the gluon distribution to 50$\%$ and its shape to much better accuracy.  Clearly the shape has increasing discriminatory power as $\eta$ increases.

We emphasize again that we have used the integrated PDFs obtained from the various `global' analyses at face value, simply to illustrate that $\bb$ data at the high values of $\eta$, accessible at LHCb, offer a powerful probe of the gluon distribution at small $x$, {\it provided} that the cuts given in (\ref{eq:cut-t}) and (\ref{eq:cut-y})  are applied. The LHC data will be able to probe a low $x$, low $Q^2$ domain well beyond the range of the data fitted in the `global' PDF analyses. Of course, in this domain, we have no reason to trust any of the predictions obtained from `extrapolated' PDFs. We merely show the curves obtained from six different PDF sets to illustrate the potential ability to constrain the gluon in the $10^{-5}\lapproxeq x \lapproxeq 10^{-3}$ domain by measuring $\bb$ events at high $\eta$, with cuts (\ref{eq:cut-t}) and (\ref{eq:cut-y}) imposed.

\section{Conclusion}
At LHC energies, the $\bb$ cross section predicted within the conventional NLO collinear approach has huge uncertainties (up to a factor of 4)
arising from the variation of the factorization scale, $\mu_F$.
This uncertainty can be strongly reduced by selecting events where the transverse momenta of the two $B$-mesons balance each other to some accuracy. If the sum of the momenta $|\vec p_{1T}+\vec p_{2T}|<k_0$, then the scale 
 $\mu_F\simeq k_0$. This offers the possibility to use $\bb$ 
 production at the LHC, particularly in the LHCb experiment, to study the $x$-dependence of the gluon 
 distribution down to $x\sim 10^{-5}$  at  rather low scales, $\mu^2\sim 4$ GeV$^2$, where 
the present HERA and Tevatron data do not constrain the  behaviour of the parton densities, and where different parton analyses propose quite different gluons.

We have considered the renormalization and factorization scale dependences of the $\bb$ cross section after the cuts (\ref{eq:cut-t}) and (\ref{eq:cut-y}), 
on the transverse momenta and rapidities of the $B$-mesons, have been imposed. In this way, we have  demonstrated how such events may determine the behaviour of the gluon distribution down to $x\sim 10^{-5}$. In Fig.~\ref{fig:bb5} we showed the differences obtained using six different set of PDFs\footnote{With our cuts, a negative gluon distribution leads to a negative $\bb$ cross section and so such a PDF set is rejected in the corresponding kinematic domain. This happens, for example, for MSTW08 gluons at very low $x$ and low scales, where the PDFs have been extrapolated well below the region of the data fitted in the global analysis. Of course, in this domain a negative gluon should not be taken literally; it is a way to account for (negative) absorptive effects which become essential at very low $x$ and low scales. For the results obtained from integrated partons in Figs. \ref{fig:bb3} and \ref{fig:bb5}, the MSTW08 gluon distribution is positive throughout the relevant domain, but the predictions based on unintegrated partons do sample a small region where the gluon is negative.} {\it simply for illustration}, bearing in mind that extrapolations of existing PDFs into this domain are unreliable.

\section*{Acknowledgements}
We thank Graeme Watt for useful discussions. MGR thanks the IPPP at Durham University for hospitality. EGdeO is supported by CNPq (Brazil) under contract 201854/2009-0, and MGR is supported by the grant RFBR
11-02-00120-a, and by the Federal Program of the Russian State RSGSS-65751.2010.2.

  M.~Cacciari, M.~Greco and P.~Nason,
  JHEP {\bf 9805}, 007 (1998),
  arXiv:hep-ph/9803400;\\
  M.~Cacciari, S.~Frixione and P.~Nason,
  JHEP {\bf 0103}, 006 (2001),
  arXiv:hep-ph/0102134.

\bibitem{petr}
 C. Peterson, D. Schlatter, I. Schmitt and Peter M. Zerwas, 
Phys. Rev. {\bf D27}, 105 (1983);\\
 V.G. Kartvelishvili, A.K. Likhoded and V.A. Petrov, 
Phys. Lett. {\bf B78}, 615 (1978).
\bibitem{cn}
M. Cacciari, P. Nason,
Phys. Rev. Lett. {\bf 89}, 122003 (2002), 
arXiv:hep-ph/0204025.
\bibitem{MCFM} J.M.~Campbell, R.K.~Ellis and C.~Williams, MCFM home page, http://mcfm.fnal.gov;\\
see also, for example,  J.M.~Campbell and R.K.~Ellis,
  Phys.\ Rev.\  {\bf D60}, 113006 (1999).
  [hep-ph/9905386],
J.M.~Campbell, R.K.~Ellis and C.~Williams,
[arXiv:1105.0020 [hep-ph]].
\bibitem{nason}
 P. Nason, S. Dawson and R.K. Ellis, 
 Nucl. Phys. {\bf B327}, 49 (1989), Erratum-ibid. {\bf B335}, 260 (1990).
\bibitem{ellis}
 R.K. Ellis and J.C. Sexton,
 Nucl. Phys. {\bf B282}, 642 (1987). 
\bibitem{MRW} A.D. Martin, M.G. Ryskin and G. Watt, Eur. Phys. J {\bf C66}, 163 (2010), arXiv:0909.5592.
\bibitem{smith} 
 W. Beenakker, W.L. van Neerven, R. Meng and G.A. Schuler,
Nucl. Phys. {\bf B351}, 507 (1991).
\bibitem{jlz} H. Jung, M. Kraemer, A.V. Lipatov and N.P. Zotov,
JHEP {\bf 1101}, 085 (2011), arXiv:1009.5067.

\bibitem{PDF}
  R.D.~Ball {\it et al.},
  Nucl.\ Phys.\  B {\bf 849}, 296 (2011),
  arXiv:1101.1300;\\
  F.D.~Aaron {\it et al.} [H1 and ZEUS Collaboration],
  JHEP {\bf 1001}, 109 (2010),
  arXiv:0911.0884;\\
  M.~Gluck, P.~Jimenez-Delgado and E.~Reya,
  Eur.\ Phys.\ J.\   {\bf C53}, 355 (2008),
  arXiv:0709.0614;\\
  S.~Alekhin, J.~Blumlein, S.~Klein and S.~Moch,
  arXiv:0908.3128;
Phys. Rev. {\bf D81}, 014032 (2010), arXiv:0908.2766 [hep-ph].


\thebibliography{} 
\bibitem{LHCb} 
R. Aaij {\it et al.}  [LHCb Collaboration],
Phys. Lett. {\bf B694}, 209 (2010),
  arXiv:1009.2731 [hep-ex].




\bibitem{MSTW08} 
  A.~D.~Martin, W.~J.~Stirling, R.~S.~Thorne and G.~Watt,
  Eur.\ Phys.\ J.\  C {\bf 63}, 189 (2009),
  arXiv:0901.0002 [hep-ph].

\bibitem{CT10} 
  H.~L.~Lai, M.~Guzzi, J.~Huston, Z.~Li, P.~M.~Nadolsky, J.~Pumplin and C.~P.~Yuan, Phy. Rev. {\bf D82}, 074024 (2010), 
  arXiv:1007.2241 [hep-ph].

\bibitem{FONLL} 
\end{document}